\documentclass[12pt,preprint]{aastex}

\def\lsim{\lower 2pt \hbox{$\, \buildrel {\scriptstyle <}\over
         {\scriptstyle \sim}\,$}}
\def\lambar{\lambda\llap {--}}

\begin{document}


\title{Population synthesis of radio and $\gamma$-ray millisecond pulsars
from the Galactic disk}


\author{Sarah A. Story} 
\affil{Hope College, Department of Physics, 27 Graves Place \\ 
Holland, MI 49423-9000} 

\email{inertialtensor@gmail.com}

\author{Peter L. Gonthier} 
\affil{Hope College, Department of Physics, 27 Graves Place \\ 
Holland, MI 49423-9000} 

\email{gonthier@hope.edu}

\and 

\author{Alice K. Harding} 
\affil{NASA Goddard Space Flight Center, Laboratory for High Energy Astrophysics \\ 
Greenbelt, MD 20771}

\email{harding@twinkie.gsfc.nasa.gov}



\begin{abstract}
We present results of a population synthesis of
millisecond pulsars from the Galactic disk. Excluding globular clusters,
we model the spatial distribution of millisecond pulsars by assuming
their birth in the Galactic disk with a random kick velocity and evolve
them to the present within the Galactic potential.  We assume that
normal and millisecond pulsars are standard candles described with a
common radio luminosity model that invokes a new relationship between
radio core and cone emission suggested by recent studies.  In modeling
the radio emission beams, we explore the relativistic effects of time
delay, aberration and sweepback of the open field lines.  While these
effects are essential in understanding pulse profiles, the
phase-averaged flux is adequately described without a relativistic
model.  We use a polar cap acceleration model for the $\gamma$-ray
emission.  We present the preliminary results of our recent study and
the implications for observing millisecond pulsars with GLAST and AGILE.
\end{abstract}





\section{Introduction}

Gamma-ray pulsars are the brightest objects in the sky above 100 MeV and
the only identified GeV Galactic sources.   Although there are presently
around 1700 known radio pulsars, only a relatively small fraction of
these are detected at higher energies. The EGRET detector on the Compton
Gamma-Ray Observatory (CGRO) detected six $\gamma$-ray pulsars with high
confidence \citep{Thom04}, five of which were known radio pulsars. In
addition, there were a few $\gamma$-ray pulsars detected with lower
confidence, one of which was the millisecond PSR \object{J0218+4232}.  The Large
Area Space Telescope (GLAST), set for launch in late 2007, will have a
sensitivity about 30 times better than that of EGRET and is expected to
detect many more $\gamma$-ray pulsars.  

Population synthesis can predict the number of expected $\gamma$-ray pulsar
detections, but the results are highly model-dependent.  As a consequence
of this sensitive model dependence, the number of $\gamma$-ray
pulsars that GLAST does detect, especially the ratio of radio-loud to
radio-quiet pulsars, will be an excellent discriminator between $\gamma$-ray models.  Using polar
cap models to describe the $\gamma$-ray emission, population synthesis
studies of \citet{Gon02,Gon04} have estimated that GLAST should
detect several hundred pulsars.  Population synthesis studies for outer
gap models \citep{Jiang06,Hard06}
generally predict fewer radio-loud $\gamma$-ray pulsars and a much higher
ratio of radio-quiet to radio-loud pulsars.  These studies evolved
neutron stars in the Galaxy from birth distributions of period, magnetic
field, position and space velocity, but excluded the population of
millisecond pulsars (MSPs).  Since MSPs are thought to be
recycled through spin-up by accretion from a binary companion, their
evolution is somewhat more complicated.  

Because of their very short periods, MSPs can have
spin-down luminosities that are comparable to those of young pulsars and
therefore should be sources of high-energy photons.  Indeed, a much
higher fraction of the total radio MSP population (around
225) have been detected as X-ray sources (around 60, including Globular
Cluster pulsars) compared to normal pulsars.  The pulsed X-ray spectra
of most MSPs in the Galactic plane are dominated by non-thermal emission
\citep{Kuip03}, so these sources must be accelerating particles to high
energy.  Polar cap models predict that MSPs should produce a high-energy
emission component due to curvature radiation up to energies around 50
GeV \citep{Hard05, Lou00}.  Outer gap models also predict that MSPs
could produce high-energy emission from particles moving downward toward
the stellar surface from the outer gap \citep{Zhang03}. Predictions for
the number of $\gamma$-ray MSPs in globular clusters have been discussed
by \citet{Wang05}.

Studies of the radio characteristics of MSPs indicate
some differences with those of the normal pulsar population.
\citet{Kramer98} and also \citet{Bail97} found that MSPs are somewhat
less luminous than normal pulsars.  However, these studies compare the
pseudo-luminosities, i.e. the observed radio flux times the square of
the distance, which are dependent on solid angle and observer viewing
angle, and do not address real luminosities.  These studies also found
that the beam radii of MSPs, although larger on average than those of
normal pulsars, are smaller than expected by an extrapolation of beam
radii with a $P^{-1/2}$ dependence of the normal pulsars
population. The average MSP beam widths seem to reach a saturated
value of around $30^\circ - 50^\circ$, whereas the extrapolated value is
greater than $100^\circ$. We will argue however that relativistic
effects \citep{Hard06} may strongly influence the shape of MSP profiles.
MSP pulse profiles have considerably larger duty cycles than normal
pulsars and the beaming fractions are thought to be around  75\%
\citep{Kramer98}.  The profiles were originally thought to have more
complex structure, a feature that was attributed to the presence of
multipole field components, possibly induced by the accretion phase
\citep{Krol91, Kramer97}.  But the observations of \citet{Kramer98} do
not support this claim and show that MSP profiles display in general
only marginally more complexity.  The more observable details afforded
by the larger duty cycles make the profiles seem more complex with the large
beams increasing the chance of detecting emission from both poles.
Additionally, the emission height dependence on frequency
(radius-to-frequency mapping) is less strong, and this is understandable
since the MSP magnetospheres are much smaller. In general, MSP radio
profiles are consistent with an expanded version of normal pulsar
emission on open dipole field lines, with the smaller widths being
attributed to emission occurring only on a subset of the open field
lines or to relativistic distortions of the beam.

We present in this paper new results of a population synthesis of MSPs
born within the Galactic disk.  In this study we define MSPs as those pulsars with period derivatives $\log( \dot P )
< -19.5 -2.5\log( P )$.  This criterion includes the binary pulsars
J1744-3922 and B0655+64 with periods of 172.4 ms and 195.7 ms,
respectively, but excludes the binary pulsar J1711-4322 with a period of
102.6 ms. By means of a Monte Carlo population code, we treat them as
point particles and evolve their trajectories, periods and period
derivatives from their birth (at time of their last spin-up phase)
forward in time to the present.  We exclude MSPs in globular clusters
from our study.  After determining a present-day spatial equilibrium
distribution, we assign radio and $\gamma$-ray characteristics to each
MSP and then filter its properties for detection through a select group
of radio surveys and the $\gamma$-ray instruments EGRET, AGILE and
GLAST.  We normalize the number of simulated radio MSPs to the number
detected by the group of radio surveys, allowing us to predict the
number of radio-loud and radio-quiet $\gamma$-ray MSPs detected by
EGRET, AGILE and GLAST.  

\section{Selected Radio Surveys}
We recently added the Swinburne Intermediate Latitude survey to our
selected group of radio surveys used in previous studies.  Of the
ten surveys, six of them, Arecibo 2 \& 3, Greenbank 2 \& 3, Molongo 2
and Parkes 2, have an observing frequency around 400 MHz, and four
surveys at 1400 MHz include Parkes 1, Jodrell Bank 2, Parkes Multibeam,
and Swinburne Intermediate Latitude.  In our computer code, we use the
characteristics of these ten surveys to determine the minimum flux
threshold for each simulated pulsar (for details see \citet{Gon02,
Gon04}) at the observing frequency of the survey. While we recognize
that some of these radio surveys were not very sensitive to the
detection of MSPs, we include them to confirm null detections.
When the simulated radio flux is below the flux threshold $S_{min}$ of
all the radio surveys that potentially are viewing the simulated pulsar,
we refer to the pulsar as being radio-quiet. Otherwise, the pulsar is
radio-loud. These characterizations are relative to the sensitivity
of the surveys and are not addressing the nature of the emission
process. We then are restricted to comparing the characteristics of the
pulsars simulated by our computer code to those MSPs detected by this
select group of radio surveys, excluding MSPs detected only by other
surveys. 

\section{Kick Velocity and Equilibrium Spatial Distribution}

We assume exponential spatial distributions of the form described in the
study of \citet{Pac90}, but using a scale height parameter above the
Galactic plane of 200 pc for MSPs in contrast to 75 pc used in our
simulations for NPs.  We use a Maxwellian distribution
with a characteristic width of $\sigma_v$ = 70 km/s to describe the
supernova kick velocity at birth from the study of \citet{Hobbs05},
resulting in an average velocity of 110 km/s.  We assume a uniform birth
rate back in time to 12 Gyr.  To determine an equilibrium spatial
distribution, we evolved a group of 40,000 neutron stars to the present
without making any selections.  As in our recent studies, \citet{Gon07}
and \citet{Story06}, we use the Galactic potential (mass model 2) of
\citet{Dehnen98} (W. Dehnen private communication) to evolve the
trajectories of MSPs using a 5th order Cash-Karp Runge-Kutta routine
\citep{Press92}.  While MSPs are born in binary systems, we have not
attempted to simulate the evolution of the binary system, but rather
treat the system as a point particle and evolve it within the Galactic
potential.  With low kick velocities, the MSPs remain bound to the
Galaxy and achieve normalized equilibrium distributions in Galactic
radius (R) and Galactic height (Z) shown in Figure 1 as solid curves
with the initial distributions represented by the dotted curves.
 
Fitting the equilibrium distributions results in radial and out-of-plane
scale heights of 4.2 kpc and 0.50 kpc, respectively.  The Z scale height
of 0.50 kpc is in good agreement with the one obtained by \citet{Cord97}
of 0.50 kpc and with the recent determinations of the
scale heights of low mass X-ray binaries (LMXB) by \citet{Grimm02} of
0.41 kpc.  In subsequent calculations, we use these
equilibrium distributions to randomly select the present-day spatial
distribution of MSPs within the Galaxy.  In choosing the R and Z values
of MSPs from such an evolved distribution rather than evolving each MSP
from its birth location, we ignore any dependence of evolved R and Z
on MSP age.

 According to \citet{Shklo70}, the transverse motion of a pulsar causes
a change in the pulsar's detected period.  Due to the Doppler effect, an
increase in the pulsar's period derivative is observed.  The
contribution to $\dot P$ due to this effect is given by \citep{Man99}
\begin{equation}
\dot P = 2.43\times 10^{-27}\left({P\over {\rm ms}}\right)\left({\mu\over {\rm mas/yr}}\right)^2\left({d \over pc}\right),
\end{equation}
where $P$ is the pulsar period, $\mu$ is the proper motion, and
$d$ is the distance.
For MSPs, whose period derivatives are extremely small, this effect
can increase period derivatives significantly, in some cases by as much as 90\%
as noted by \citet{Toscano99} who have also included the smaller  
effects of Galactic rotation and vertical acceleration.  In this study,
we only include the Shklovskii effect due to the proper motion of the simulated
pulsar.

\section{Magnetic Field and Initial Period}
While it may be that the magnetic field of MSPs decays slowly, we have
assumed a constant field during their lifetime.  However, due to
selection effects, the observed field distribution will differ from the
initial distribution. We explored power laws with various indices to
characterize the distribution of pulsar surface magnetic field at the
birth line.  While \citet{Cord97} preferred a power law with an index of
-2, the group of 22 MSPs used in their study had smaller magnetic
fields.  For the group of 56 MSPs in this study, we find a preferred
index of -1 with a normalized distribution of the form
\begin{equation}
n(B_8 ) = {1 \over { {B_8 \ln\left( {B_{\rm max}\over {B_{\rm min}} }\right) }}},
\end{equation}
where $B_8$ is in units of $10^8$ G.  We choose a $B_{\rm min}=1$ and
a $B_{\rm max}=10^4$ that
provides the best agreement between various simulated and detected
distributions.  

Having the magnetic field for a simulated MSP, we can obtain its initial
period with a spin-up relation.  \citet{Alpar82} and similarly
\citet{Bhat91} found that for recycled pulsars spun up by accretion at
the Eddington limit from a binary companion, the initial period is
estimated by the expression
\begin{equation}
P_{\rm{o}}  = 0.3B_8^{6/7} {\rm{ms}},
\end{equation}
which can be rewritten as a birth line in the $\dot P - P$ diagram by the form
\begin{equation}
\log \dot P_{\rm{o}}  = \frac{4}{3}\log P_{\rm{o}}  - 15.4,
\end{equation}
where the period is in seconds and the period derivative is ${\rm
s\cdot{s^{-1}}}$. Having the magnetic field, the initial period, and the
age, we can determine the present period and period derivative of the
MSP assuming a pure dipole spin down.  However, recent studies of LMXBs
where the spin periods of accreting neutron stars were measured have
allowed the estimation of the surface magnetic fields.  \citet{Lamb05}
conclude that the general properties of LMXBs can be understood if the
accretion rates range from the Eddington critical accretion
rate to $5\times 10^{-4}$ times that rate.  These different accretion rates
lead to different birth lines in the $\dot P-P$ diagram. To incorporate a
distribution of accretion rates, we have included an approximate
procedure in which we dither the intercept of the following birth line
\begin{equation}
\log \dot P_{\rm{o}}  = \frac{4}{3}\log P_{\rm{o}}  - 14.9 - \delta ,
\end{equation}
where the dithering parameter $\delta$ varies from 0 to 2.8.  The birth line
can be reformulated by the expression
\begin{equation}
P_{\rm{o}}  = 0.18 \times 10^{3\delta /7} B_8^{6/7} {\rm{ms}}.
\end{equation}
The extremes of the dithering parameter $\delta$ represent the approximate
Eddington accretion critical rate and $5\times 10^{-4}$ times that rate (see figure 4
in \citet{Lamb05}).   While we explored a Gaussian distribution of the
dithering parameter, we obtain better agreement with a ramp
distribution that increases by a factor of 4 between 0 and 2.8.  With the magnetic field randomly
selected from the distribution of Equation 2, we then use Equation 6 to
obtain the initial period of the simulated MSP.  However, we impose a
minimum initial period of $P_{o_{\rm min}}$ = 1.3 ms in accord with the RXTE studies
of LMXBs by \citet{Chak05}.  Having randomly determined the
pulsar's age assuming a uniform birth rate, we spin the pulsar down to
obtain the present period and period derivative assuming constant
magnetic surface field.  A constant magnetic field used in this study of
MSPs is in contrast to the assumption of magnetic field decay in our
simulations of NPs \citep{Gon02, Gon04, Gon07}.

\section{Radio Luminosity and Beam Geometry}

In our previous studies of NPs \citep{Gon02, Gon04, Gon07}, we adopted
the intrinsic radio luminosity model given by the general form
\begin{equation}
L = L_o P^{ \alpha} \dot P^{\beta},
\end{equation}
from \citet{Arz02} (ACC) with $L_o=2.1\times 10^{12} {\rm{ mJy}} \cdot {\rm{kpc}}^{\rm{2}}  \cdot {\rm{MHz}}$,
$\alpha = -1.3$ and $\beta = 0.4$, but reduced the
luminosity $L_o$ by a factor of 60 \citep{Gon04} and by 73 \citep{Gon07}
to obtain adequate agreement between the simulated birth rate, flux, and
distance distributions and those detected, particularly by the Parkes
Multibeam Pulsar survey (PMBPS) (see \citet{Gon04}).  In ACC, the
predicted birth rate was 0.13 neutron stars per century.  In order to
obtain a birth rate near 2 neutron stars per century, we had to decrease
the luminosity constant $L_o$. The assumed radio luminosity is directly
related to the simulated neutron star birth rate. Various studies
propose neutron star birth rates that span somewhat of a range. For
example, the a range of birth rates from 0.9 to 1.9 per century was
obtained from the analysis of PMBPS by \citet{Vran04}.  In a recent
population synthesis, \citet{Fauch06} estimate a larger birth rate of
about 2.8 per century, whereas \citet{Lorimer06} find $1.4 \pm 0.2$ per
century from their analysis.  More recently \citet{Gon07} determined the
reduction factor of the radio luminosity by normalizing the simulated
neutron star birth rate to rate of Type II supernovae of 2.1 per
century, following the work of \citet{Tamm94}, using only the Parkes
Multibeam survey, as we may have the best description of the flux
threshold $S_{min}$ for that survey (Crawford, private communication).

While the neutron star birth rate of 2.1 per century provides a
constraint on the luminosity of NPs, MSPs appear in a very different
region in the $\dot P-P$ diagram, requiring careful consideration of the
period and period derivative dependence of the radio luminosity used by
the population synthesis study. Often, radio astronomers refer to the
radio luminosity in terms of a ``pseudoluminosity" ($Sd^2$) determined
from the radio flux $S$ observed at a specific frequency and the pulsar
distance $d$.  One cannot generally determine the intrinsic radio
luminosity of a given pulsar from the $Sd^2$, even at a given frequency,
as the detected average flux is strongly dependent on the viewing
geometry.  However, if the number of detected pulsars in a group is
large, one might be justified in assuming that the emission region of
the group of pulsars is completely sampled in a random fashion.
Therefore, the detected $Sd^2$ would be proportional to the intrinsic
luminosity of the group of pulsars if the surveys represented an
unbiased sample of the true flux distribution. However, since radio
surveys necessarily sample the high end of the flux distribution,
inferring the intrinsic luminosities of pulsars is difficult at best. In
addition, selection effects of the radio surveys may be different for
NPs and MSPs, and these effects do influence the sampling of the
emission region.  For example, the width of the pulse profile is used in
the determination of the flux threshold $S_{\rm min}$ for each radio
survey; therefore, one would expect that narrow pulse profiles or larger
inclination angles are preferably detected.  Regardless of these
effects, such comparisons are important in characterizing differences
between NPs and MSPs. \citet{Kramer98} compared the average $log(Sd^2)\
{\rm (mJy\cdot kpc^2 )}$ of a group of MSPs with that of a group of NPs
and found that  the
mean $\log( Sd^2 )$ values for their groups of 31 MSPs and of 369 NPs
of $0.5 \pm 0.2$ and $1.50\pm0.04$, respectively.  The authors conclude
that MSPs and NPs have very different radio luminosity distributions.
Since this study was performed, the NE2001 distance model \citep{Cord02}
was developed.  We use the pulsar distance given by the NE2001 model
unless the distance has been determined by other more accurate methods.
After this distance correction and using the 1400 MHz fluxes in the ATNF
catalogue that have a recorded uncertainty \citep{Lorimer95,
Kramer98, Lorimer06, Manchester01, Hobbs04, Kramer03, Faulkner05,
Faulkner04,Morris02, Stairs05}, we find average $\log( Sd^2 )$ values of
$0.38 \pm 0.15$ for our select group of 39 MSPs (24 with S1400 fluxes)
and $1.17 \pm 0.02$ for a group of 1102 NPs (869 with S1400 fluxes
(references in ATNF catalogue)) detected by Parkes Multibeam and the
Swinburne Intermediate Latitude surveys.  The uncertainties are just the
statistical error in the mean and do not take into account the error in
the fluxes and distances.  Subtle selection effects may be introduced
within the context of a population statistics study  with a group
of surveys that have different sensitivities. As a result, it
may be inappropriate to compare these mean $\log( Sd^2 )$ values with
those from \citet{Kramer98} because of the different samples of pulsars.
The MSPs recently discovered by the Parkes Multibeam and Swinburne
surveys are not only more distant  but also weaker with effectively
smaller $\log( Sd^2 )$. We have chosen to reproduce the $\log( Sd^2 )$ values of
the normal pulsars and MSPs detected by these surveys to constrain our
luminosity model.

We adjust our intrinsic radio luminosity model such that the simulated
mean $\log( Sd^2 )$ values approximately match both the mean values of $0.38 \pm 0.15$
for MSPs and $1.17 \pm 0.02$ for NPs detected by both the Parkes
Multibeam and the Swinburne Intermediate Latitude surveys. We find that
the exponents of the period ($\alpha$) and period derivative ($\beta$)
dependence of the luminosity are not uniquely constrained to specific
values but are correlated with each other.  We find better
agreement between the various simulated distributions for NPs with
values of $\alpha=-1.05$ and $\beta=0.37$. The coefficient of the
luminosity is adjusted to give a birth rate of 2.1 neutron stars per
century using the Parkes Multibeam and the Swinburne Intermediate
Latitude surveys.

We were able to find a common radio luminosity that provides adequate
results for both NPs and MSPs, given by the expression
\begin{equation}
L = 1.76 \times 10^{10} P^{ - 1.05} \dot P^{0.37} {\rm{ mJy}} \cdot {\rm{kpc}}^{\rm{2}}  \cdot {\rm{MHz}}.
\end{equation}

In this study, we assume that the radio emission originates from a
single core beam and a single cone beam axially aligned with the
magnetic field from each pole of the star and isotropically distributed
in the sky.  As indicated by \citet{Kramer98}, the profiles of MSPs
display similar structure to those of NPs, suggesting that the core or
central beam and the conal beam make similar contributions to NP and MSP
profiles.  However, as discussed later, due to the rapid rotation of
MSPs, relativistic effects are more evident in the pulse profiles of
MSPs than in those of NPs \citep{Xilo98}.  We use the same description
for the core beam as in our past study of \citet{Gon04}, but we replace
the geometry of the cone beam from the \citet{Mitra99} with the
description of \citet{Kijak98, Kijak03} having the form
\begin{equation}
\rho _{{\rm{cone}}}  = 1.24^ \circ  r_{{\rm{KG}}}^{1/2} P^{ - 1/2},
\end{equation}
which describes the characteristic width of the conal beam at 0.1\% of
the peak intensity of the profile assuming that the edges of the pulse
are along the last open field line.  They find that the altitude of the
emission $r_{KG}$ in stellar radii is given by
\begin{equation}
r_{{\rm{KG}}}  = 40\nu _{{\rm{GHz}}}^{{\rm{ - 0}}{\rm{.26}}} \dot P_{ - 15}^{0.07} P^{0.3},
\end{equation}
where $\dot P_{-15}$ is in units of $10^{-15}{\rm \ s\cdot{s^{-1}}}$.  In
this formulation, the overall
period dependence of the characteristic width of the conal beam is $\rho_{\rm cone}\sim
P^{-0.35}$, reproducing well the trend observed for NPs and MSPs \citep{Kramer98, Kramer02}.
	 
In the formulation of a Gaussian conal beam, the characteristic width
$\rho_{\rm cone}$ requires the specification of the radius $\bar \theta$
and width $w_e$ of the annulus of the beam. The characteristic width of
\citet{Kijak98,Kijak03} is for 0.1\% of the peak intensity making
$\rho_{\rm cone}=\bar\theta+2.63w_e$.  Within such a model, one is free
to specify some form for either $\bar\theta$ or $w_{e}$.  We
adopted $w_e=0.18 \rho_{\rm cone}$ in our study of both normal and MSPs.  We find
that the population statistics are not very sensitive to the choice of
this value, which we have arrived at from our study of three peak pulse
profiles of radio pulsars. In our study of the characteristics of core
and cone beams of about 20 radio pulsars whose profiles have three
peaks, we found (\citet{Gon06} and in preparation) that the ratio of
core-to-cone peak fluxes does not follow the $P^{-1}$ dependence
suggested in the work of ACC.  In the ACC model, the profiles of MSPs
are dominated by the central core beam, so that even at small impact
angles the conal beam would not be seen.  As seen in the study of
\citet{Kramer98}, similar complexity is manifested in the MSPs profiles
as in profiles NPs, indicating that the conal beam is quite prominent. 
Recent polarimetry studies of young radio pulsars also call the ACC
core-to-cone ratio into question. \citet{Craw01} and \citet{Craw03}
measured the radio polarization of nine pulsars, six of which had
characteristic ages less than 100 kyr. They found that the profiles of
all the young pulsars indicate a significant degree of linear
polarization and a low degree of circular polarization, which is a
traditional feature of conal emission. In this study, the two older
pulsars with characteristic ages greater than 1 Myr do not show any
degree of polarization.  They conclude, as did \citet{Man96}, that the
profiles of young short period pulsars tend to be characterized by
partial cones.  Similarly, in a more recent study \citet{John06}
measured polarization profiles of 14 young pulsars with ages $<$ 75 kyr.
They found that the profiles are dominated by linear polarization,
suggesting that the core beam is weakly manifested, if at all.  Because
of these inconsistencies, we have reassessed the relationship between
the core and cone beams (\citet{Gon06}), using three-peak normal pulsars
and a few three-peak MSPs.  In our three-peak pulsar study, we find for
the ratio of peak fluxes a power-law dependence with a break in period
at about 0.7 s having the approximate form
\begin{equation}
r = \left\{ \begin{array}{l}
 25P^{1.3} \nu _{{\rm{GHz}}}^{{\rm{ - 0}}{\rm{.9}}} ,{\rm{    }}P < 0.7{\rm{s}} \\ 
 4P^{ - 1.8} \nu _{{\rm{GHz}}}^{{\rm{ - 0}}{\rm{.9}}} ,{\rm{    }}P > 0.7{\rm{s}} .\\ 
 \end{array} \right.
\end{equation}

Hence in this model, short period pulsars are much less core dominated
than in the ACC model.  The model does not distinguish between young
short-period pulsars and MSPs, although there is evidence
(\citet{Kramer99}) that the ratio of core-to-cone peak fluxes may be
different for a young pulsar and a MSP with the same period.  The shapes
of MSP profiles are complicated by relativistic effects, but due to the
wide emission beams, contributions from both poles are often seen in the
profiles, making the analysis of the viewing geometry and conal widths
difficult.  With this model, we do accommodate more complex pulse
profiles for short period pulsars than in the ACC model.  

In describing the core and cone angle-integrated spectra, we assume
indices of -2.36 and -1.72, respectively. While these indices are not
measurable, they are related in a complex fashion, due to selection
effects, to the measured spectral indices.  Using these values, our
simulation gives a flux-averaged spectral index of $-1.80\pm 0.05$ for
MSPs. Within our select group of surveys, we find 20 MSPs within the
ATNF catalogue whose spectral indices have recorded uncertainties
\citep{Toscano98,Lorimer95} with a mean of $-1.89\pm 0.10$.
\citet{Kramer98} found a mean spectral index of $-1.8\pm 0.1$ for a
group of 32 MSPs.  They also point out that MSPs have significantly
steeper mean spectral indices than NPs.  For a group of 346 NPs, they
find a mean spectral index of $-1.60\pm 0.04$.  In our simulation, we
found it necessary to have softer angle-integrated spectral indices for
the core and cone beams of NPs, resulting in a mean spectral index of
$-1.58\pm 0.01$.  This value agrees well with the mean spectral index of
$-1.66\pm 0.04$ for the 245 NPs with indices that have recorded
uncertainties in the ATNF catalogue (references therein).

Adopting this radio luminosity and beam geometry model, we attempt to
describe radio emission properties from both NPs and MSPs
\citep{Gon07,Hard06}.  The success of this radio model is
remarkable in describing the physical luminosities of both NPs and
MSPs, given the range of three decades in period and seven decades in
period derivative between NPs and MSPs. However, it is clear that there
are many complexities that may not modeled correctly and that the population
synthesis study may not be very sensitive to various characteristics of the beam geometry.

\section{Relativistic Effects in the Radio Profiles} Because of their
rapid rotation, MSPs can display very different features in their pulse
profiles than those observed in the profiles of NPs.  These features can
be partly understood in terms of the relativistic effects of aberration,
time delay and the sweep back of the retarded dipole field
\citep{Dyks04a,Dyks04b}.  The shorter the pulsar period, the more
dramatically these effects are manifested in the profiles. According to
Equation 10, the conal emission of pulsars with short periods occurs at
high altitude relative to the light cylinder radius.   For MSPs the
altitude of emission for the cone beams ($r_{KG}$) is  $ \sim 0.2 - 0.6\
R_{LC}$.  The combined effects of time delay and aberration result in a
shift in phase of the conal beam.  Deformations in the shape of some
profiles of MSPs are a result of the distortion of the open field volume
of the retarded dipole \citep{Dyks04a}. While relativistic effects are
important in understanding the shape of the profile and in studying
correlation with high-energy profiles, in a recent study
\citep{Story06}, we find that the phase-averaged flux is very similar
for relativistic and nonrelativistic emission.  Since in population
statistics studies, the average flux of the profile is the determining
parameter that is compared to the threshold of each radio survey, we
have neglected these effects in this study.

\section{Gamma-ray Luminosity and Beam Geometry}
Because of the lack of a theoretical understanding of the mechanism that
produces the radio emission, the radio emission model is
phenomenological by nature.  On the other hand, the $\gamma$-ray emission models are
motivated by theory.  Two competing $\gamma$-ray emission models, the polar cap
and outer gap models, assume different locations in the magnetosphere
where the acceleration of charge occurs.  In the outer gap model, the
acceleration of charge takes place in vacuum gaps formed along the last
open field line, between the null surface and the light cylinder.  While
we are beginning to incorporate an outer gap in our simulations
\citep{Hard06}, treating it on the same footing as
the polar cap model using the same group of evolved neutron stars and
selection criteria, we use only the polar cap model in this study of
MSPs. Recent outer gap models for MSPs \citep{Zhang03} actually
assume that the high-energy emission originates not from the outer gap
but near the neutron star, from particles flowing down from the gaps,
and thus would have a very different high-energy emission geometry from
that of normal pulsars.

Two regions in the $\dot P-P$  diagram, separated by the curvature radiation (CR)
pair death line, differentiate the $\gamma$-ray luminosities and beam
geometries within the polar cap model.  Pulsars above the CR line can
produce CR photons that are energetic enough to pair produce in the
strong magnetic field above the neutron star surface.  The CR-initiated pair
cascades have sufficient multiplicity to fully screen the parallel
electric field everywhere except in the slot gap along a narrow region
near the last open field line where the primary particles are
accelerated along unscreened electric field lines.  Recent studies
\citep{Musl03,Musl04} of the emission above the CR pair death line in
the polar cap model have provided a framework for understanding
$\gamma$-ray emission from the slot gap.  The electrons in the low
altitude slot gap initiate pair cascade emission, which was incorporated
in our simulations of \citet{Gon04,Gon07}.  In addition, \citet{Musl04}
find that the primary particles along the last open field lines can be
accelerated to high altitudes as the potential in the slot gap remains
unscreened and the emission beam forms a characteristic caustic
component \citep{Dyks03}.  In recent population studies of NPs, we have
begun to include both the low and high altitude emission geometries
\citep{Gon07,Hard06} and both are included in this study for those simulated
MSPs above the CR line.

However, as one can see in Figure 2, all the MSPs with the exception of
one lie below the CR line where curvature radiation no longer produces
pairs.  These pulsars can produce much weaker pair cascades through
inverse Compton scattering radiation. When pulsars are no longer able to
produce abundant pair cascades, they do not form slot gaps above the
polar cap \cite{Musl03} as the parallel electric field is
not screened on all open magnetic field lines.  As the $\gamma$-ray energies
are below the pair threshold, the curvature radiation from the primary
electrons escapes out to infinity along all open field lines. In this
pair-starved polar cap model, the primary particles continue to
accelerate and radiate to high altitudes above the polar cap out to the
light cylinder along all the open field lines.  However, CR
 emitted by the particles  can be inhibited by resonant
cyclotron absorption of radio emission \citep{Hard05}
where synchrotron radiation is more efficient particle energy loss mechanism than CR.  Since synchrotron
photons usually have lower energies than curvature photons, we have not
included this mechanism in the pair-starved model in this study.  The
CR emission model below the CR pair
death line used in this study is developed from the work by \citet{Hard05}.

The inclination angle $\alpha$ and the viewing angle $\zeta$ determine which open
field lines are sampled by the line of sight.  As illustrated in Figure
3, the line of sight intersects tangentially a particular open field
line with a radius of curvature $\rho_c$ at a radial distance r and polar
angle $\theta$.  The open field lines are defined within the polar cap angle
given by the expression 
\begin{equation}
\theta _{pc}  = \sin ^{ - 1} \left( {\frac{{2\pi R}}{{cP}}} \right)^{1/2} ,
\end{equation}
where P is the pulsar period in seconds, $R$
is the neutron star radius ($10^6$ cm)  and c is the speed of light.  We
partition the open field lines through a dimensionless parameter, $\xi$,
that varies between 0 and 1 and is defined as 
\begin{equation}
\xi  = \frac{{\theta _s }}{{\theta _{pc} }}{\rm{, }}
\end{equation}
where $\theta_s$ is the polar
angle of the open field line at the intersection with the surface of the
star.  The particle emits curvature photons along the line of sight at
an angle $\theta_\gamma\ =\ {3\theta / 2}$ given by
\begin{equation}
\cos \theta _\gamma   = \sin \alpha \sin \zeta \cos \phi  + \cos \alpha \cos \zeta ,
\end{equation}
where $\phi$ is the phase angle, $\alpha$ is the magnetic inclination and $\zeta$ is the
viewing angle.

We approximate the accelerating electric field from Equation 14 of
\citet{Hard98} using the expression
\begin{equation}
E_{||}  = \sin ^4 (\theta _{pc} )B_{12} \left[ {\kappa \cos \alpha \eta ^{ - 4}  
+ \frac{1}{8}\sin ^{1/2} (\theta _{pc} )\xi \sin \alpha \cos \varphi 
\eta ^{ - 1/2} } \right](1 - \xi ^2 ){\rm{\ statvolt}} \cdot {\rm{cm}}^{{\rm{ - 1}}} ,
\end{equation}
where $\kappa$ = 0.15 is the general relativistic inertial frame dragging
factor, $B_{12}$ is the surface magnetic field in units of $10^{12}$ G,  $\eta$ =r/R
is the dimensionless radius and $\varphi$ is the azimuth angle around the
magnetic pole given by the expression
\begin{equation}
\cos \varphi  = \frac{{\cos \zeta  - \cos \alpha \cos \theta _\gamma  }}
{{\sin \alpha \sin \theta _\gamma  }}.
\end{equation}

The gain in energy of the accelerating primary electron is compensated
by the CR losses \citep{Hard02,Bulik00,Lou00},
so that the electron Lorentz factor $\gamma$
becomes radiation-reaction limited to
\begin{equation}
e\left| {E_{||} } \right| \sim \frac{{2e^2 \gamma ^4 }}{{3\rho _c^2 }},
\end{equation}
where $\rho_c$ is the classical radius of curvature of the field line given by
\begin{equation}
\rho _c  = \frac{{r\left( {1 + \cos ^2 \theta } \right)^{3/2} }}{{3\sin 
\theta (1 + \cos \theta )}}.
\end{equation}
At low altitudes near the stellar surface, $r\ <\ R(\sin\theta_{pc}/3 +1)$,  the near
surface electric field is approximated from equation 21 in \citet{Hard01}
by the expression 
\begin{equation}
E_{||}  =  - 4.05B_{12} \sin ^3 \theta _{pc} (\eta  - 1)\left[ {0.15(1 - 
\xi ^{2.19} )^{0.705} \cos \alpha  + \frac{3}{8}\sin \theta _{pc} \xi ^{1.015} (1 - 
\xi ^2 )^{0.65} \sin \alpha \cos \varphi } \right]
\end{equation}
and goes to zero at the stellar
surface.  The CR spectrum therefore has a spectral
index of -2/3 with a high-energy cutoff $\epsilon_{CR}$ given by
\begin{equation}
\varepsilon _{CR}  = \left( {\frac{{3\lambar_C\gamma ^3 }}{{2\rho_c }}} \right) = 
\hbar c\left( {\frac{3}{2}} \right)^{7/4} \left( {\frac{{E_{||} }}{e}} 
\right)^{3/4} \rho _c^{1/2} {\rm{,}}
\end{equation}
in $m_ec^2$ units where $e$ is the
electron charge and $\lambar_C\equiv\hbar/(m_e c)$ is the electron Compton wavelength.

The instantaneous curvature radiation power spectrum has the form
\citep{Jack75} 
\begin{equation}
P_{CR} (\varepsilon _\gamma  ,\theta _\gamma  ,\xi ) = 3^{1/2} \alpha _{\rm fine} 
\gamma \left( {\frac{c}{{2\pi \rho _c }}} \right)\kappa \left( {
\frac{{\varepsilon _\gamma  }}{{\varepsilon _{CR} }}} \right),
\end{equation}
where $\epsilon$ is the photon energy, $\gamma$ is the
particle Lorentz factor, $\alpha_{\rm fine}$ is the fine structure constant, and the
$\kappa(x)$ function is defined as
\begin{equation}
\kappa (x) = x\int_x^\infty  {K_{5/3} (x')dx'}  \approx \left\{ \begin{array}{l}
 2^{2/3} \Gamma \left( {\frac{2}{3}} \right)x^{1/3} ,x \ll 1, \\ 
 1.253x^{1/2} e^{ - x} ,x \gg 1 \\ 
 \end{array} \right.
\end{equation}
We find that the low energy asymptotic form provides a sufficiently
accurate description of the power spectrum above $\epsilon_\gamma$=100 MeV and the
instantaneous curvature photon radiation spectrum per primary is given by
\begin{equation}
N_{CR} (\varepsilon ,\theta _\gamma  ,\xi ,\varphi ) = \frac{{3^{1/6} 
\Gamma \left( {\frac{2}
{3}} \right)\alpha _{fine} }}
{{\pi \hbar ^{1/3} }}\left( {\frac{c}
{{\rho _c }}} \right)^{2/3} \varepsilon ^{ - 2/3} \sim\frac{{0.518\alpha _{fine} }}
{{\hbar ^{1/3} }}\left( {\frac{c}
{{\rho _c }}} \right)^{2/3} \varepsilon ^{ - 2/3} ,
\end{equation}
which is similar to equation 7 in \citet{Hard05}. This expression
can be integrated from $\epsilon_\gamma$=100 MeV to the high-energy cut off 
$\epsilon_{CR}$ to give
the expression
\begin{equation}
N_{CR} ( > \varepsilon _\gamma  ,\theta _\gamma  ,\xi ,\varphi ) = 
\frac{{1.554\alpha _{fine} }}{{\hbar ^{1/3} }}\left( {\frac{c}{{\rho _c }}} 
\right)^{2/3} \left( {\varepsilon _\gamma ^{1/3}  - \varepsilon _{CR}^{1/3} } 
\right),
\end{equation}
which provides the number of curvature photons emitted along the line of
sight $\theta_\gamma$ from an open field line $\xi$ at a radial distance $r$ per primary
particle, where $r$ is given by the expression
\begin{equation}
 r = \frac{{R\sin ^2 \theta }}{{\sin ^2 (\xi \theta _{pc} )}}.
 \end{equation}
The Goldreich-Julian current of primary particles from the
polar cap is uniformly distributed over the polar cap and is given by
the expression
\begin{equation}
\dot N_p  = 1.3 \times 10^{30} B_{12} P^{ - 2} {\rm\ particles/s}.
\end{equation}
The number of primary particles in a particular patch on the stellar
surface at R, $\theta_s$, and $\varphi$, which gives the number of primaries along a
particular field line $\xi$, is provided by the expression
\begin{equation}
\dot n_{GJ}  = \frac{{\dot N_{GJ} }}{{2\pi (1 - \cos \theta _{pc} )}}.
\end{equation}
The total photon luminosity from one pole can be obtained by integrating
over the open field volume to give
\begin{equation}
L_{{\rm{total}}} ( > \varepsilon _\gamma  ) = \dot n_{GJ} \int_0^{2\pi } 
{\int_0^1 {\int_R^{R_{LC} } {N_{CR} ( > \varepsilon _\gamma  ,\theta _\gamma  ,
\xi ,\varphi )\theta _{pc} \sin \xi \theta _{pc} drd\xi d\varphi } } }.
\end{equation}
With 
\begin{equation}
 \left( {\frac{{dr}}{{d\theta }}} \right) = \frac{{2R\sin \theta \cos \theta }}
 {{\sin ^2 (\xi \theta _{pc} )}}, 
 \end{equation}
the integral over $r$ can be written as an integral over $\theta$ to give
\begin{equation}
L_{{\rm{total}}} ( > \varepsilon _\gamma  ) = \int_0^{2\pi } {\int_0^1 
{\int_{\theta (R)}^{\theta (R_{LC} )} {\left( {\frac{{2R\dot n_{GJ} N_{CR} 
( > \varepsilon _\gamma  ,\theta _\gamma  ,\xi ,\varphi )\theta _{pc} \cos 
\theta }}{{c\sin (\xi \theta _{pc} )}}} \right)\sin \theta d\theta d\xi d
\varphi } } } .
\end{equation}
where the integral over $\xi$ (the $\theta_{pc}$ has been included in the integrand)
is over the open field lines and $\xi\theta_{pc}$ is the angle at the footpoint of the
field line on the stellar surface.
The expression in parentheses represents the photon emission from a
particular field line at $\xi$ and $\eta$, integrated above $\epsilon_\gamma$=100 MeV, along
the line of sight into a solid angle $d\Omega$ and is given by the expression
\begin{equation}
\frac{{dN_{CR} ( > \varepsilon _\gamma  ,\theta _\gamma  ,\xi ,\varphi )}}
{{d\Omega }} = \frac{{1.552\alpha _{fine} \dot N_{GJ} cR\theta _{pc} 
\rho _c^{ - 2/3} \cos \theta \left( {\varepsilon _\gamma ^{1/3}  - \varepsilon 
_{CR}^{1/3} } \right)}}{{\pi (1 - \cos \theta _{pc} )(\hbar c)^{1/3} \sin (\xi 
\theta _{pc} )}}.
\end{equation}
The total emission from the polar cap for a particular phase bin $\phi$ is
obtained by integrating over parameter $\xi$ to include the contributions of
all the open field lines along a line of sight defined by the viewing
angle $\zeta$.  The total emission in the phase bin is given by
\begin{equation}
N_{CR}^{tot} ( > \varepsilon _\gamma  ,\phi ) = \int_0^1 {\frac{{dN_{CR} ( > 
\varepsilon _\gamma  ,\theta _\gamma  ,\xi ,\varphi )}}{{d\Omega }}d\xi } ,
\end{equation}
Having calculated the $\gamma$-ray pulse profile for each MSP for a given
viewing geometry, we average the profile to obtain the average photon
flux, and compare it to the instrument threshold.  Our resulting $\gamma$-ray
efficiency (7\%) for the total spectrum and profiles for the nearby MSP 
PSR \object{J0437+4715} are in
agreement with those obtained by \citet{Vent05}.

\section{Gamma-ray All Sky Threshold Maps}

From the simulated $\gamma$-ray pulse profile, we obtain an average flux
that we compare to all sky threshold maps for EGRET, AGILE and GLAST
Large Area Telescope (LAT).  We use the recently revised EGRET map
that includes the dark clouds \citep{Casand07}, which has led to
a radical reassessment of the EGRET unidentified sources.  The GLAST
threshold has been improved and updated (Grenier \&
Casandjian private communication) as a 1 year GLAST LAT threshold map. 
The all sky map for AGILE (Pellizzoni private communication) has not
been recently updated in our computer code.  The detection of radio and
$\gamma$-ray point sources within the code are independent of each
other, allowing the tagging of radio-quiet (below the survey flux
thresholds) and radio-loud $\gamma$-ray MSPs.

\section{Results}
To improve the simulated statistics, we run the simulation to obtain ten
times the number of detected MSPs and then normalize the distributions
accordingly.  In Figure 4, we present Aitoff projections for the
detected (a) and simulated (b) MSPs.  Since MSPs are closer to us and
are much older than NPs, the graphs indicate larger out-of-plane
distributions than those of NPs.

Of the ten radio surveys that we have included in our simulation, the
surveys most sensitive to MSP detection are Parkes 2, Parkes
Multibeam and Swinburne Intermediate Latitude which together account for
majority of the 56 MSPs, as seen in Table 1. Within
the limited statistics, there is good agreement between the number of
MSPs detected and simulated among the ten surveys used in our selected
group.  The few detections in radio surveys that are not very sensitive
to the detection of MSPs are also well reproduced by the simulation.

The distributions in the $\dot P-P$ diagram are compared
in Figure 5 for detected (a) and simulated (b) MSPs. The dotted broken
lines represent the pair death lines for curvature radiation (CR) and
for nonresonant inverse Compton scattering (NRICS).  Below the CR death
line, CR no longer is able to produce pairs.  However, a limited number
of pairs can still be produced via inverse Compton scattering above the
NRICS death line.  Below the NRICS death line, pair production is no longer
possible, presumably inhibiting radio emission mechanisms. The upper and lower MSP birth lines
indicated by the solid lines represent the approximate Eddington
critical accretion rate (upper line) and   $5\times 10 ^{-4}$ times that rate (lower
line) discussed earlier.  With the $B^{-3/2}$ magnetic field
distribution and the uniform distribution of birth lines between the
upper and lower MSP birth lines, we reproduce the observed broadness of
the pulsar distribution in the $\dot P-P$ diagram.  

In Figure 6, we present histograms of the indicated measured and derived
pulsar characteristics for the detected (shaded histograms), and the
simulated (unshaded histograms) MSPs.  Although the poor statistics make
it difficult to judge the agreement, the simulation appears to reproduce
the observed distributions fairly well.  The clump of detected MSPs
observed in Figure 5a is exhibited as narrow peaks in the period and
period derivative distributions.  The narrow peak in the period
derivative distribution is not present in the simulated distribution. 
In the comparisons of the age distributions, we calculated the
traditional characteristic age $P/(2\dot P)$  for the simulated pulsars.
 In the comparisons of the distances and dispersion measures of the
MSPs, we have used the new distance model of \citet{Cord02}.  The
distance of the detected pulsar is established typically by its measured
dispersion measure and location in the sky. The new distance model is
used to determine the simulated dispersion measure from the simulated
distance and location in the sky for each MSPs, which is then used in
the calculation of the flux threshold $S_{min}$ for each radio survey. 
Both the simulated dispersion measure and distance distributions
indicate good agreement with the corresponding distributions of detected
MSPs with a few more simulated closer pulsars. The simulated radial and
height distributions within the Galaxy are consistent with those
distributions of the detected pulsars.  However, the simulated spectral
index distribution is narrower than that of the detected pulsars. For
the pulsar characteristics in Figure 6, we have performed a
Kolmogorov-Smirnov (KS) statistics test of the histograms assuming a
significance level ($\alpha$) of 5\%. We indicate in Figure 6 the
resulting p-values for each of the comparisons. Sometimes the D
statistic and the critical value are quoted in the literature and
compared to each other to decide on the null hypothesis. The critical
value is derived from $\alpha$ and the number of samples.  Equivalent
comparisons can be made between the p-value and $\alpha$.  If the
p-value statistic is larger than $\alpha$, the null hypothesis is not
excluded at the level of significance determined by $\alpha$.  Our
results indicate that in all comparisons, the distributions of simulated
pulsars are consistent with those detected with the exception of the
distributions of the Galactic height z and S400 whose p-value is less
that 0.005. In the case of the S400 distributions, the statistics are
insufficient in the number of measured flux densities at 400 MHz for an
adequate comparison.  We simulate slightly more pulsars with larger
S1400 fluxes. We applied the two dimensional KS test to the unbinned
period and period derivative distributions using the prescription of
\citet{Press02} and obtained a p-value of only of 0.03 at the same significance
level of 5\% indicating that the detected and simulated $\dot P- P$
distributions are not entirely consistent with each other. While the KS tests might
be helpful in such comparisons, the numerous assumptions within the
simulation, such as the radio luminosity, the pulsar distance, and birth
distributions, contain systematic uncertainties. Therefore, we do not
want to overemphasize the importance of these statistical tests.

The detection of a radio pulsar in the computer code is independent of
the detection of a $\gamma$-ray pulsar, allowing the identification of
radio-quiet and radio-loud $\gamma$-ray pulsars.  By radio-quiet, we mean that
the radio flux of the simulated pulsars is below the flux threshold of
the surveys that could potentially observe it; otherwise, the pulsar is
flagged as radio-loud.  In Figure 7, we show the simulated radio-quiet
and radio-loud $\gamma$-ray MSPs for the instruments EGRET (upper right), AGILE
(lower left) and GLAST LAT (lower right).  For comparison, we also show
the $\gamma$-ray pulsars that were detected by EGRET (upper left), including
the MSP \object{J0218+4232}. In the case of Geminga (cross in upper right), it is
not clear whether its radio silence results from a misalignment of
the $\gamma$-ray and radio beams or from its intrinsic radio weakness.
While it is unclear how many of the EGRET unidentified sources might be
MSPs detected as point sources, the simulation predicts that a few may
be radio-quiet $\gamma$-ray MSPs.

All the simulated $\gamma$-ray pulsars that are detected by EGRET, AGILE and
GLAST lie below the curvature radiation pair death line in the $\dot P-P$
diagram (Figure 7).  As discussed above, in
this pair-starved polar cap model, curvature radiation escapes along all
open field lines.  However, as indicated by \citet{Hard05},
curvature radiation can be limited when the accelerated
particles undergo resonant cyclotron absorption of radio photons under
special circumstances. The excited particles emit synchrotron radiation
typically at lower energies than curvature radiation.  While we are in
the process of developing a full three-dimensional study of these
processes, in the present calculation we have not limited the curvature
radiation in altitude. Therefore, our results may be somewhat
optimistic.  In Figure 7, the $\gamma$-ray MSPs cluster in the lower left of
the $\dot P-P$ diagram because the $\gamma$-ray luminosity increases with decreasing $P$.

In Table 2, we present the $\gamma$-ray MSPs statistics for radio-loud
and radio-quiet pulsars.  Due to the uncertainties in the radio luminosity, 
we vary the coefficient of the radio luminosity 
in Equation 8 by $\pm 10\%$ to obtain a range in radio-quiet $\gamma$-ray pulsars.
Since the simulation is normalized to the detected number of radio pulsars, there is
no variation in the number of radio-loud $\gamma$-ray pulsars.  
Of the 33-40 radio-quiet $\gamma$-ray MSPs
predicted to be detected by GLAST, about 6 have fluxes above the
$10^{-7}$ photons/$\rm cm^2$/s level (Grenier, private communication),
that GLAST may be able to detect through blind period searches. 
However, since 70\% of the MSPs in the ATNF catalogue are in binaries,
many of these radio-quiet $\gamma$-ray pulsars are expected to be in
binary systems.  It will be difficult for GLAST to detect pulsed
emission from these MSPs in blind period searches due to timing
uncertainties associated with the orbital motion.  Without radio
ephemerides, GLAST will detect binary MSPs only as point sources.

In order to explore the nature of the radio-quietness of the GLAST
pulsars, we present in Figure 8 histograms of the radio flux at 1400 MHz
of the detected (solid dark gray histograms) and simulated (open thin
histograms) radio pulsars in (a) and the radio flux of GLAST
radio-quiet (solid light gray histograms) and radio-loud (open thick
histograms) in (b).  As expected, GLAST radio-loud pulsars have large radio
fluxes.  However, the GLAST radio-quiet pulsars form two groups, one
with a flux distribution very similar to the radio detected and
simulated pulsars in Figure 8a and the other with a much lower flux
distribution that would be difficult to detect without deep radio
pointings.

The Parkes Multibeam and Swinburne Intermediate Latitude surveys are the
two most sensitive surveys in our select group of radio surveys. 
However, they cover a rather narrow strip along the Galactic disk, and
MSPs seem to be rather scattered in space and are not as correlated with
the Galactic disk as are NPs.  As a result, many of the GLAST detected
pulsars are radio-quiet because they happen to be outside the sensitive
survey regions or have radio fluxes that are slightly lower than the
survey thresholds.

In Figure 9, we plot the impact angle $\beta=\zeta-\alpha$ as a function of the radio flux
at 1400 MHz for the simulated group of GLAST radio-quiet pulsars.  We
can see that the two groups of fluxes correspond to two groups of impact
angles.  The MSPs in the low flux group with large impact angles will not be
observed as radio sources without deep radio observations.  However, the simulation
suggests that they can be seen as $\gamma$-ray point sources.
In the pair-starved model of $\gamma$-ray emission, curvature
radiation occurs along all open field lines all the way out to the light
cylinder.  While the intensity of curvature radiation decreases at
higher altitudes for nearby pulsars, we predict that GLAST should be
able to detect them as point sources, yet their radio emission will be very
difficult to detect as the radio beam and the $\gamma$-ray beam are 
separated by large angles.  However, the higher flux group with smaller impact
angles would be detected by GLAST as point sources and potentially are
detectable with pointed radio observations.

As a result, our simulations suggest that it will be important to follow
up GLAST detections with radio observations in order to discover a
greater number of radio-loud $\gamma$-ray pulsars. The predicted energy spectra
of MSPs are very different than those of either NPs or AGNs. In this
pair-starved polar cap model, the energy cutoffs of the MSP $\gamma$-ray
spectra are predicted to be in the 10-50 GeV energy region \citep{Hard05}
suitable for GLAST.

Clearly the number of detected radio-quiet and radio-loud $\gamma$-ray MSPs
depends strongly on the detection thresholds of radio surveys and $\gamma$-ray
instruments.  In Figure 10, we present the Log(N)-Log(S) for the radio
flux at 400 MHz (a) and 1400 MHz (b) and for EGRET (c) and GLAST (d). 
The light shaded histograms represent the simulated undetected source
counts.  The dark histograms represent the simulated detected source
counts by the select group of ten radio surveys at 400 MHz (a) and 1400
MHz (b).  For EGRET (c) and GLAST (d), the dark and medium dark
histograms represent the simulated source counts for radio-loud and
radio-quiet $\gamma$-ray MSPs, respectively.

The appearance of fractions of a pulsar in the figure is a result of the
re-normalizing the distributions because the simulation was run for ten
times the number of detected MSPs to improve the simulated statistics. 
Assuming a uniform distribution of sources in space with isotropic
emission of equal flux, one expects to a slope in the Log(N)-Log(S)
diagram of -3/2.  Indeed the tails of these distributions have slopes
very similar to -3/2.  While for EGRET and GLAST the Log(N)s of
simulated detections have a sharp decrease with decreasing flux, the
Log(N)s of simulated radio detections have a more gentle roll over,
which may reflect the larger number of factors that determine the
minimum flux threshold for the radio surveys.   Both the number of
undetected radio and $\gamma$-ray sources follow a simple power law far below
the sensitivity of the radio surveys and $\gamma$-ray instruments.

\section{Discussion}
We have presented a Monte-Carlo simulation of the Galactic-plane
population of MSPs and corresponding predictions for the numbers of both
radio-quiet and radio-loud MSPs that are detectable as $\gamma$-ray
sources. In order to accomplish this, we made significant improvements
in our population statistics Monte Carlo code by adding the more
realistic description of the Galactic potential of \citet{Dehnen98} with
a more accurate 5th order Cash-Karp Runge-Kutta trajectory integration
routine \citep{Press92}, using improved all-sky threshold maps for EGRET
\citep{Casand07} and GLAST (Grenier \& Casandjian, private
communication), and including the Swinburne Intermediate Latitude radio
survey \citep{Edwards01}.  We have modified the intrinsic radio
luminosity model of ACC to describe the total radio luminosity of both
normal and MSPs.  From our study of radio pulsars with three-peak
profiles \citep{Gon06,Gon07}, we find that the ratio of the core-to-cone
peak fluxes for short period pulsars is much smaller for short period
pulsars than the ratio predicted by ACC model. Although MSPs have on
average larger pulse widths than NPs, this resulting radio beam geometry
model provides a reasonable agreement to the average pseudoluminosities
and pulse widths at 1400 MHz and pulse widths of both NPs and MSPs. We include a
$\gamma$-ray beam geometry and luminosity model of the curvature
radiation from all open field lines within a pair-starved polar cap
model \citep{Hard05} that accounts for $\gamma$-ray emission below the
curvature radiation pair death line, which is applicable to MSPs.  These
improvements in our computer code have allowed us to study the
population statistics of radio and $\gamma$-ray MSPs to complement our
previous studies of NPs from the Galactic disk.  As a first order study,
we believe that we have improved the simulation by including a beam
geometry for MSPs and an intrinsic radio luminosity whose functional
form allows for the description of the population statistics of normal
pulsars.  The actual beam geometry of MSPs may turn out to be
significantly different and require a different model than the one
proposed here from the work of \citet{Kijak98,Kijak03}.  The population
statistics study is not sensitive enough to discriminate between beam
geometry models as the average flux of the profile is the quantity that
is compared to the $S_{min}$s of the radio surveys.  A detailed study
including relativistic effects will be necessary along with high quality
polarization data may enable establishment of systematic trends that
allows for the development of an adequate beam model for MSPs.

In this study, we do not attempt to describe the numerous MSPs within
globular clusters.  Limiting our study to the MSPs born in the Galactic
disk, we treat them as point particles, evolving their location within
the Galactic potential and their spin-down, after spin-up by mass
accretion from a binary companion star has ended.  This group of old,
short period pulsars with low magnetic fields are given radio and
$\gamma$-ray beam and luminosity characteristics and filtered through a
set of ten radio surveys and the $\gamma$-ray instruments EGRET, AGILE
and GLAST.  Our simulation is run until the number of simulated radio
pulsars is equal to the number detected by the same group of radio
surveys.  The radio luminosity is adjusted in our study of NPs
\citep{Gon07,Hard06} to give a birth rate of Type II supernovae of 2.1
per century to agree with the study of \citet{Tamm94} and to reproduce
the detected mean pseudo-luminosities and spectral indices of MSPs and
NPs, after which there are no further adjustments performed in the
simulation of MSPs.  Allowing for a $\pm$ 10\% variation in the radio
luminosity, we predict a birth rate of $4-5 \times 10^{-4}$ MSPs per
century, which is in good agreement with the studies of \cite{Lorimer05}
($3\times 10^{-4}$ per century) and \citet{Ferr06} ($3.2\times 10^{-4}$
per century).  While the population statistics studies of LMXBs
contain numerous uncertainties, our simulated birth
rate agrees with the estimate of the birth rate of LMXBs from
the preferred model (F) of \citet{Kiel06} ($6.5\times 10^{-4}$ per
century).  We are also in agreement with \citet{Kramer98} as our
description of the spectral properties of the core and cone beams for
MSPs requires steeper angle-integrated spectral indices than those for
the simulated NPs.  

In outer gap accelerator models \citep{Cheng86}, MSPs are
not expected to produce $\gamma$-ray emission above 1-2 GeV (\citet{Zhang03},
Hirotani private communication 2005) due to the smaller electric field and
larger radius of curvature in the gap.  Thus, detected emission from
MSPs above 5 GeV would imply an origin near the polar cap.  \citet{Zhang03}
propose that particles accelerated in the outer gaps could
be the origin of MSP radiation at super-GeV energies, but only if the
cascade emission from the particles as they travel down toward the
neutron star surface is visible.  However, it seems unlikely that both
the upward-going radio emission and downward-going $\gamma$-ray emission could
be visible to the same observer.  In that event, one would expect
the radio and high-energy peaks to be out of phase, and X-ray and radio
peaks are often in phase for MSPs (e.g.  \object{B1821-24}, \object{J0437-4715},
\object{B1921+37}).  In any case, GLAST observations of MSPs will be an excellent
model discriminator.

Our simulations predict the number of radio-loud and radio-quiet $\gamma$-ray
pulsars observed by the instruments EGRET, AGILE and GLAST.  EGRET was
only able to detect one MSP, PSR \object{J0218+4232}, during its nine-year
mission.  EGRET did not have the capability of performing blind period
searches of pulsars and required reliable radio ephemerides to detect
$\gamma$-ray pulsations.  However, these observations were prior to the Parkes
Multibeam Pulsar Survey and the Swinburne Intermediate Latitude Survey
that tripled the number of detected radio MSPs in Galaxy 
(not counting those in globular clusters).  We predict
that EGRET should have detected as point sources eight radio-loud and $\sim 20$
radio-quiet $\gamma$-ray MSPs, some of which could be associated with EGRET
unidentified sources.  However, the recently revised all-sky EGRET
threshold map with dark clouds \citep{Casand07}, used in our
simulations, has significantly affected the catalogue of EGRET
unidentified sources.  As the all-sky threshold map for AGILE has not
been recently updated, our predictions for AGILE might be somewhat optimistic. 
However, the trend in the statistics for EGRET, AGILE and GLAST suggests
that the number of radio-quiet and radio-loud $\gamma$-ray MSPs predicted to
be detected by AGILE might be reasonable.   The Shklovskii effect is a significant, positive contribution
to the intinsic $\dot P$ of the pulsar resulting in much larger measured $\dot P$s
and needs to be taken into account in population statistics studies.  The effect of having
to decrease the intrinsic $\dot P$ to accomodate this effect results is pulsars with lower
$\gamma$-ray luminosities.  We predict $\sim 38$ of radio-quiet $\gamma$-ray pulsars for GLAST with the ratio of
radio-quiet to radio-loud increasing from 1.5 for EGRET to 3.2 for GLAST.
Curvature radiation of $\gamma$-rays below the curvature pair death line occurs
along all open field lines all the way out to the light cylinder, as the
weak production of pairs through nonresonant inverse Compton scattering
is insufficient to screen the electric field.  The increased sensitivity
of GLAST results in a large number of detections at higher altitudes
where the $\gamma$-ray beam is at large angles relative to the radio beam,
resulting in radio fluxes below the flux thresholds of our select group
of radio surveys.  However, in many cases, the radio fluxes are
relatively large but are outside of the sky region of the Parkes
Multibeam and Swinburne Intermediate Latitude surveys.  MSPs are much
less confined to the plane of the Galactic disk than are NPs and
therefore often appear at high latitudes, far above the narrow belt
covered by these more sensitive surveys.  Our simulations suggest that
GLAST will detect many MSPs as point sources, and only a handful as
radio-loud pulsars.  The point sources can be identified as pulsar
candidates by their spectra.  \citet{Hard05} indicate
that the curvature radiation energy spectra of MSPs are very hard, with
cutoffs in the region of 10 to 50 GeV, well within the GLAST range. 
Such detected candidates can easily be followed up with pointed radio
observations to obtain the periods of the pulsars.  Thus, GLAST with
radio follow-up observations offers the potential to discover many new
MSPs in the Galaxy.

\acknowledgments

 We are very grateful to Isabelle Grenier and Jean-Marc Casandjian
for providing a GLAST-LAT 1 year point source threshold map.  We express
our appreciation to the anonymous referee for the insightful comments,
which led to a much-improved manuscript.  We are indebted to Dunc
Lorimer who pointed out our neglect of the Shklovskii effect in an
earlier version of the manuscript. We express our gratitude for the
generous support of the National Science Foundation (REU and
AST-0307365), the Michigan Space Grant Consortium and the NASA
Astrophysics Theory Program.

\clearpage

\begin{table}
\begin{center}
\caption{Detected and Simulated MSPs in each Radio Survey}
\begin{tabular}{cccc}
\tableline\tableline
Survey	& Frequency (MHz)	& Detected	& Simulated \\
\tableline
Arecibo 3	& 430	& 4	& 8 \\
Arecibo 2	& 430	& 2	& 1 \\
Greenbank 3	& 390	& 0	& 3 \\
Greenbank 2	& 390	& 1	& 0 \\
Molongo 2	& 408	& 0	& 1 \\
Parkes 2	& 436	& 18	& 23 \\
Parkes 1	& 1520	& 0	& 0 \\
Jodrell Bank 2	& 1400	& 0	& 0 \\
Parkes MB	& 1374	& 28	& 20 \\
Swinburne IL	& 1374	& 12	& 9 \\ 
\tableline
\end{tabular}
\end{center}
\end{table}

\clearpage

\begin{table}
\begin{center}
\caption{Simulated $\gamma$-ray MSP statistics}
\begin{tabular}{cccc}
\tableline\tableline
Instrument	& Radio-loud	& Radio-quiet	& Ratio RQ/RL \\
\tableline
EGRET detected	& 1		& ?			& \  \\
EGRET simulated	& 4		& 5 - 6		& 1.5 \\
AGILE simulated	& 7	& 11 - 13		& 1.9 \\
GLAST simulated	& 12	& 33 - 40	& 3.2 \\
\tableline
\end{tabular}
\end{center}
\end{table}

\clearpage

\begin{figure}
\epsscale{.80}
\plotone{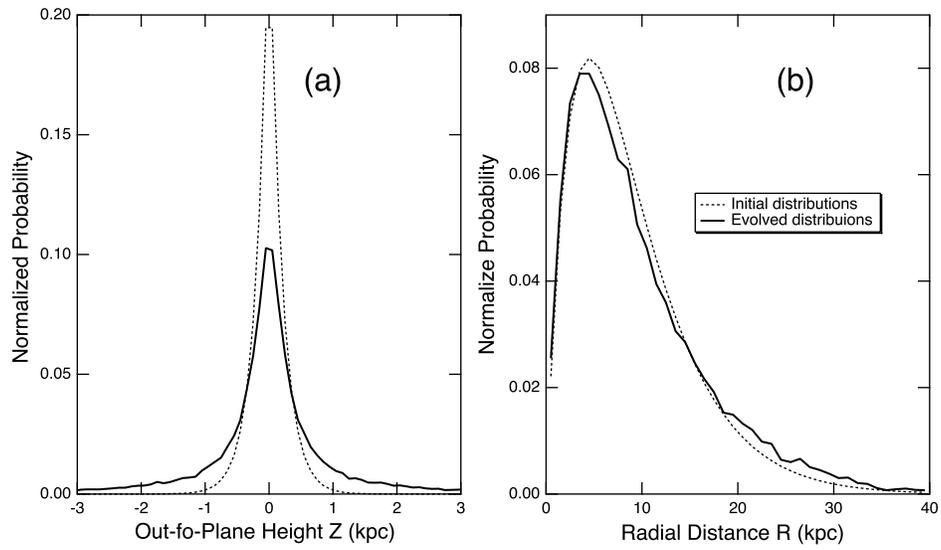}
\caption{Height above the Galactic plane Z (a) and Galactic radius R (b) 
distributions of millisecond pulsars.  Initial distributions are indicated by 
dotted curves and equilibrium distributions by solid curves.\label{fig1}}
\end{figure}

\clearpage

\begin{figure}
\epsscale{.8}
\plotone{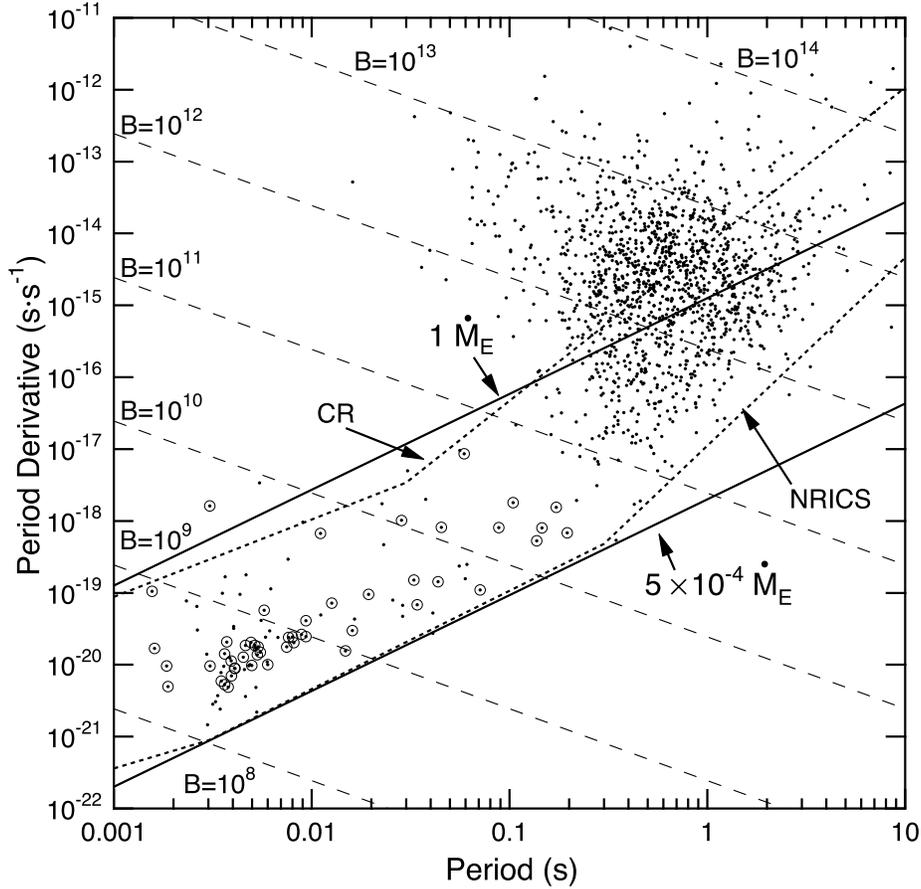}
\caption{$\dot{P}-P$ diagram of radio pulsars (dots) in the ATNF catalogue
(http://www.atnf.csiro.au/people/pulsar/psrcat/ and \citet{Man05}).  Millisecond 
pulsars detected by the group of ten radio surveys used in this study are 
indicated with an additional open circle. The thin dashed lines represent 
the indicated traditional magnetic surface strength, assuming a constant 
dipole spin-down field.  The dotted broken lines represent the pair death 
lines for curvature radiation (CR) and for nonresonant inverse Compton 
scattering (NRICS).  The upper and lower MSPs birth 
lines for Eddington critical accretion rate $\dot{M}_E$ (above) and 
$5\times 10^{-4} \dot{M}_E$  of that rate 
(below) discussed in the text are indicated by solid lines.\label{fig2}}
\end{figure}

\clearpage

\begin{figure}
\epsscale{.80}
\plotone{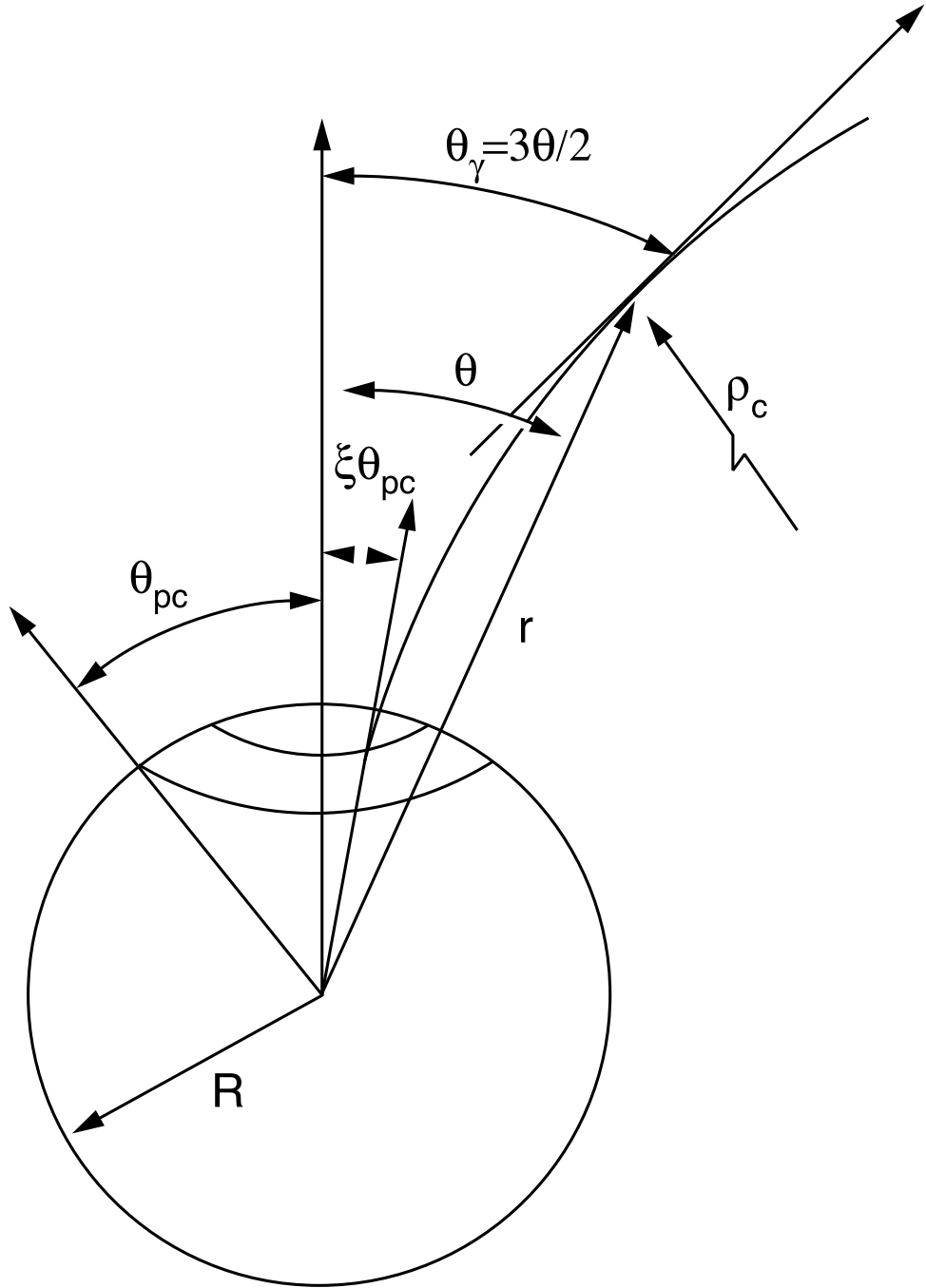}
\caption{Geometry defining the angular quantities associated with a
charged particle along an open field line defined by $\xi$ emitting
curvature radiation at $\theta$ and r along the line of sight, $\theta_\gamma$.
\label{fig3}}
\end{figure}

\clearpage

\begin{figure}
\epsscale{.80}
\plotone{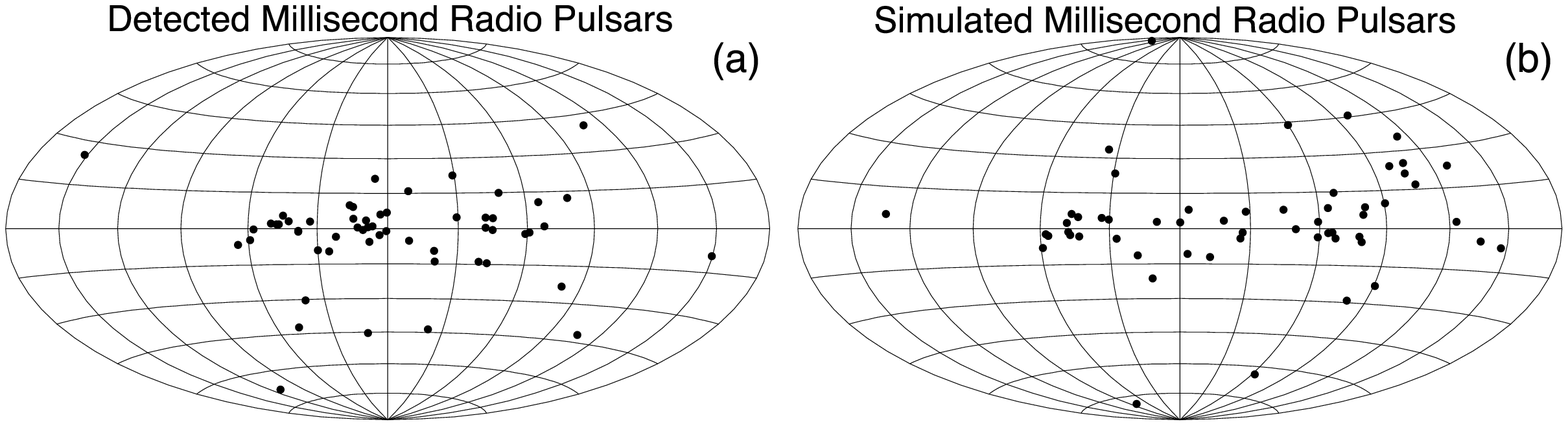}
\caption{Aitoff projections of detected (a) and simulated (b) MSPs. 
\label{fig4}}
\end{figure}

\clearpage

\begin{figure}
\epsscale{.80}
\plotone{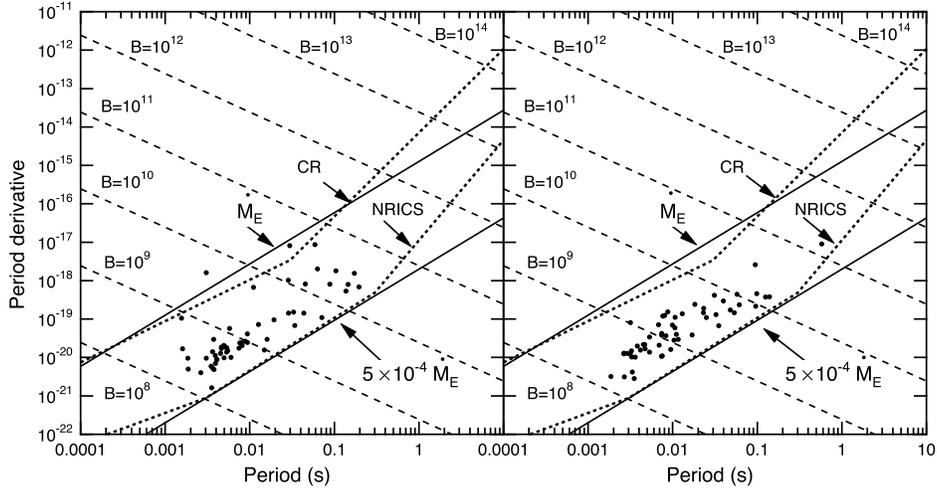}
\caption{$\dot{P}-P$ diagram of millisecond pulsars detected (a) and
simulated (b) by the group of ten radio surveys. The thin dashed lines
represent the indicated traditional magnetic surface strength, assuming
a constant dipole spin-down field.  The dotted broken lines represent
the pair death lines for curvature radiation (CR) and for nonresonant
inverse Compton scattering (NRICS).  The upper
and lower MSPs birth lines for Eddington critical accretion rate
$\dot{M}_E$ (above) and $5\times 10^{-4}$ $\dot{M}_E$ of that rate (below)
discussed in the text are indicated by solid lines.
\label{fig5}}
\end{figure}

\clearpage

\begin{figure}
\epsscale{.80}
\plotone{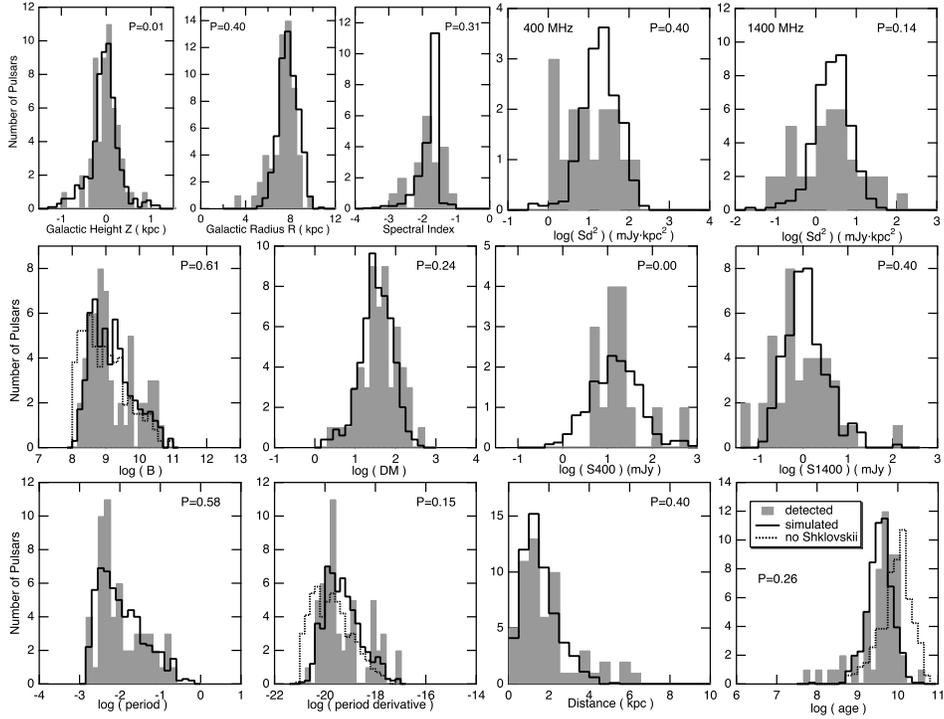}
\caption{Distributions of various characteristics indicated for detected
(shaded histograms) and simulated (unshaded histograms) MSPs from the
Galactic disk.  Also indicated is the p-value of the
Kolmogorov-Smirnov test of the binned detected and simulated sample distributions
at a significance level of $\alpha=5\%$.  The dotted histograms represent the
simulated distributions without the Shklovskii effect.
\label{fig6}}
\end{figure}

\clearpage

\begin{figure}
\epsscale{.80}
\plotone{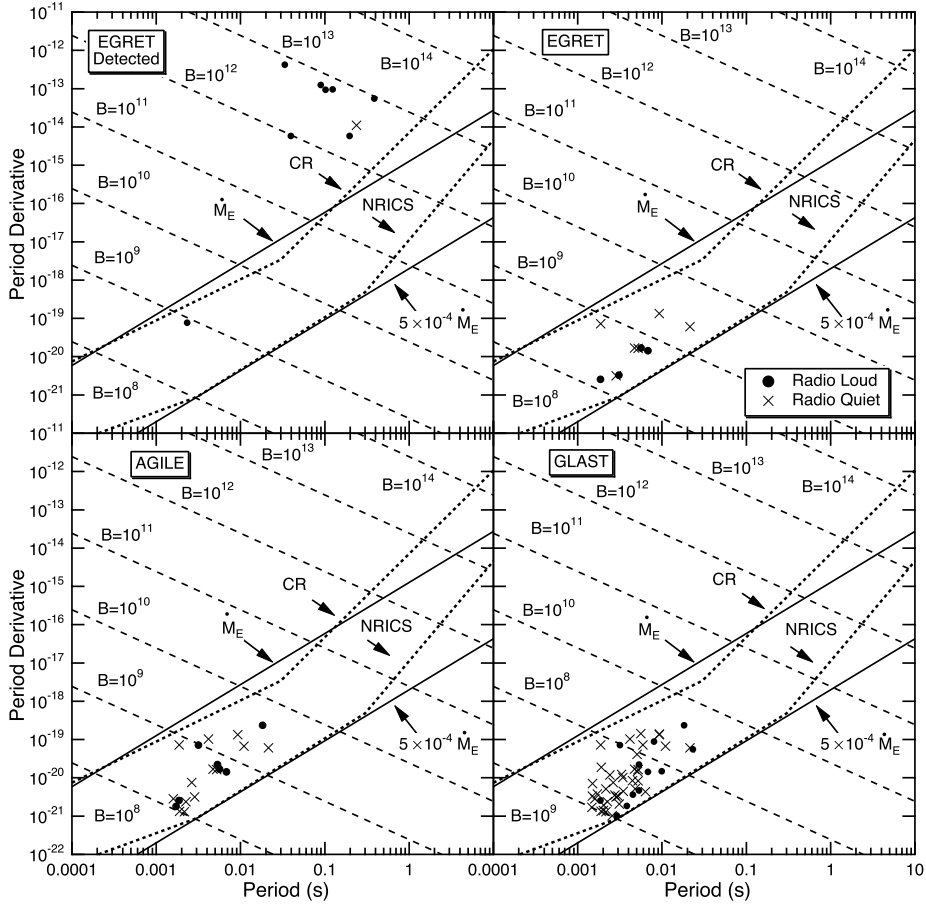}
\caption{$\dot{P}-P$ diagram of radio-quiet (crosses) and radio-loud
(solid circles) $\gamma$-ray normal and millisecond pulsars detected by
EGRET (upper left) and millisecond pulsars predicted to be detected by
EGRET (upper right), AGILE (lower left) and GLAST (lower right). The
thin dashed lines represent the indicated traditional magnetic surface
strength, assuming a constant dipole spin-down field.  The dotted broken
lines represent the pair death lines for curvature radiation (CR) and
for nonresonant inverse Compton scattering (NRICS).  The upper and lower MSPs birth lines for Eddington critical
accretion rate $\dot{M}_E$ (above) and $5\times 10^{-4}\ \dot{M}_E$ of that
rate (below) discussed in the text are indicated by solid lines.
\label{fig7}}
\end{figure}

\clearpage

\begin{figure}
\epsscale{.60}
\plotone{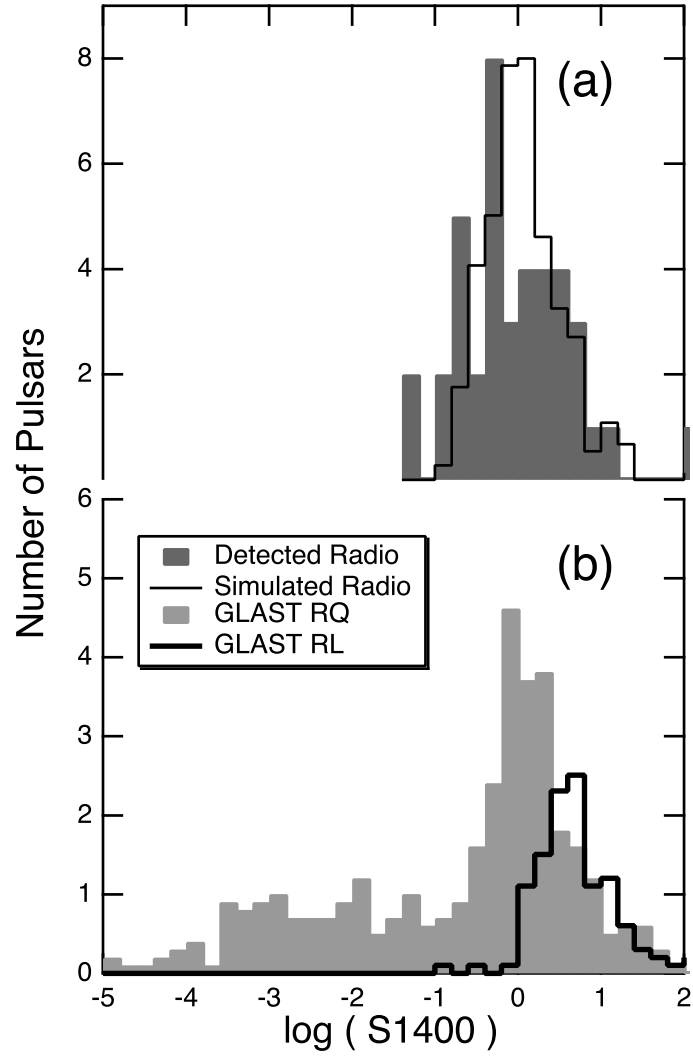}
\caption{Radio flux at 1400 MHz plotted for detected and simulated radio
pulsars in (a) and for GLAST radio-quiet and radio-loud $\gamma$-ray
pulsars in (b).
\label{fig8}}
\end{figure}

\clearpage

\begin{figure}
\epsscale{.60}
\plotone{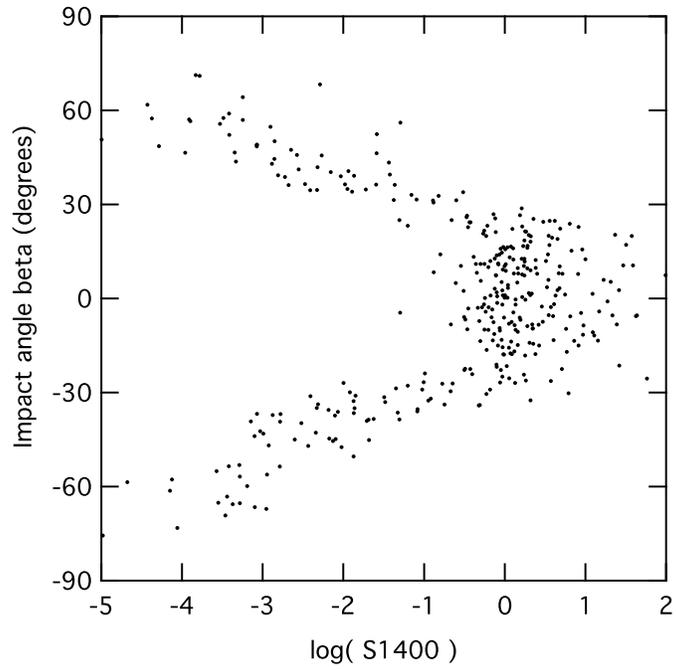}
\caption{The impact angle $\beta$ plotted as a function of the radio flux at
1400 MHz for the simulated GLAST radio-quiet $\gamma$-ray pulsars (10 times).
\label{fig9}}
\end{figure}

\clearpage

\begin{figure}
\epsscale{.60}
\plotone{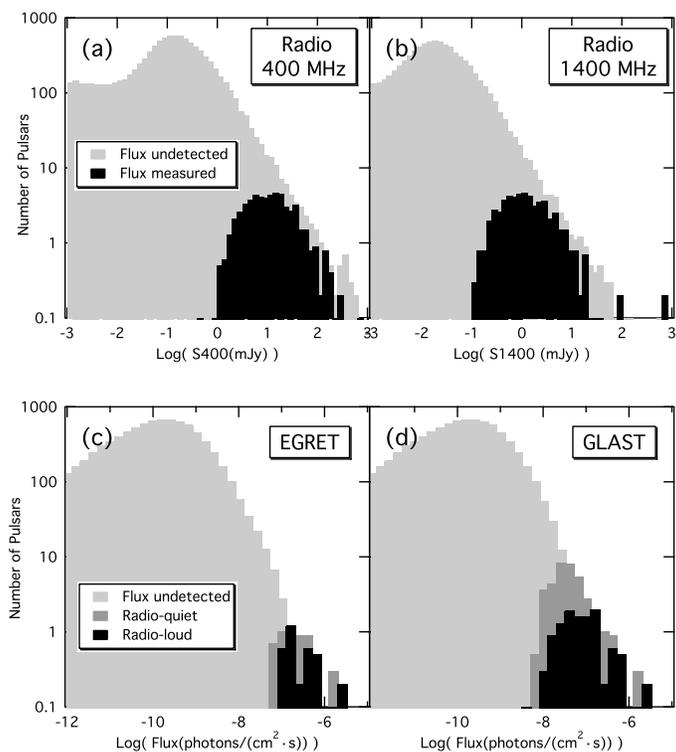}
\caption{Log(N)-Log(S) for the radio flux at 400 MHz (a), 1400 MHz (b)
and for EGRET (c) and GLAST (d).  The light histograms represent the
simulated undetected source counts.  For the radio flux, the dark
histograms indicated the simulated detected flux at 400 MHz (a) and 1400
MHz (b).  For EGRET (c) and GLAST (d), the predicted $\gamma$-ray source
counts are indicated for radio-loud (dark histograms) and radio-quiet
(medium dark histograms) $\gamma$-ray millisecond pulsars.
\label{fig10}}
\end{figure}

\end{document}